\documentclass[sigconf]{acmart}
\AtBeginDocument{%
  }

\setcopyright{acmlicensed}
\copyrightyear{2018}
\acmYear{2018}
\acmDOI{XXXXXXX.XXXXXXX}
\acmConference[Conference acronym 'XX]{Make sure to enter the correct
  conference title from your rights confirmation email}{June 03--05,
  2018}{Woodstock, NY}
\acmISBN{978-1-4503-XXXX-X/2018/06}

\acmSubmissionID{1319}

\usepackage{caption}
\usepackage{subcaption}

\renewcommand{\textcolor}[2]{#2}
\begin{document}

\title[Balancing Memorability and Efficiency in VR Authentication]{\textcolor{red}{Context or Digits? Balancing Memorability and Efficiency in Virtual Reality Authentication}}

\author{Yuxuan Huang}
\email{yhuang94@ncsu.edu}
\orcid{0000-0001-6308-6516}
\affiliation{%
  \institution{North Carolina State University}
  \city{Raleigh}
  \state{North Carolina}
  \country{USA}
}

\author{Qiao Jin}
\email{qjin4@ncsu.edu}
\orcid{0000-0001-5493-1343}
\affiliation{%
  \institution{North Carolina State University}
  \city{Raleigh}
  \state{North Carolina}
  \country{USA}
}

\author{Tongyu Nie}
\email{nie00035@umn.edu}
\orcid{0000-0003-4186-8749}
\affiliation{%
  \institution{University of Minnesota}
  \city{Minneapolis}
  \state{Minnesota}
  \country{USA}
}

\author{Victoria Interrante}
\email{interran@umn.edu}
\orcid{0000-0002-3313-6663}
\affiliation{%
  \institution{University of Minnesota}
  \city{Minneapolis}
  \state{Minnesota}
  \country{USA}
}

\author{Evan Suma Rosenberg}
\email{suma@umn.edu}
\orcid{0000-0002-4826-4561}
\affiliation{%
  \institution{University of Minnesota}
  \city{Minneapolis}
  \state{Minnesota}
  \country{USA}
}

\renewcommand{\shortauthors}{Huang et al.}

\begin{abstract}
% Motivation
%Knowledge-based authentication in Virtual Reality (VR) faces a fundamental challenge on the tradeoff between efficiency, memorability, and security. 
%While PIN-based methods enable fast password entry, memorability may suffer with strong passwords. Methods leveraging environmental cues allow for better recall but sacrifice efficiency. 
We present Adaptive Direction-Based Authentication (ADBA), a knowledge-based authentication method for Virtual Reality that decouples users' needs temporally by enforcing password creation based on virtual environment context while supporting both context- and digit-based entries during authentication. This design prioritizes memorability for new passwords and offers both efficient and memorable options to support users' evolving needs. 
We conducted a remote longitudinal study with 66 participants comparing ADBA against 6-digit PINs over 2-3 weeks. 
The results demonstrated that ADBA achieved superior memorability and lower perceived task load. \textcolor{red}{Interestingly, no participant chose to enter via digits in the study, yet they still perceived ADBA to be highly efficiency despite longer objective entry times.} ADBA also provided security benefits through randomly-generated digit representations, though some degree of password homogeneity was observed in specific virtual environments. 
Our findings suggest that ADBA offers solid advantages to the traditional PINs, and successfully addresses the tradeoffs between efficiency, memorability, and security under the usage scenarios considered in the study.
\end{abstract}



\begin{teaserfigure}
  \centering
  \includegraphics[width=0.9\linewidth, trim = 0px 5px 0px 0px, clip]{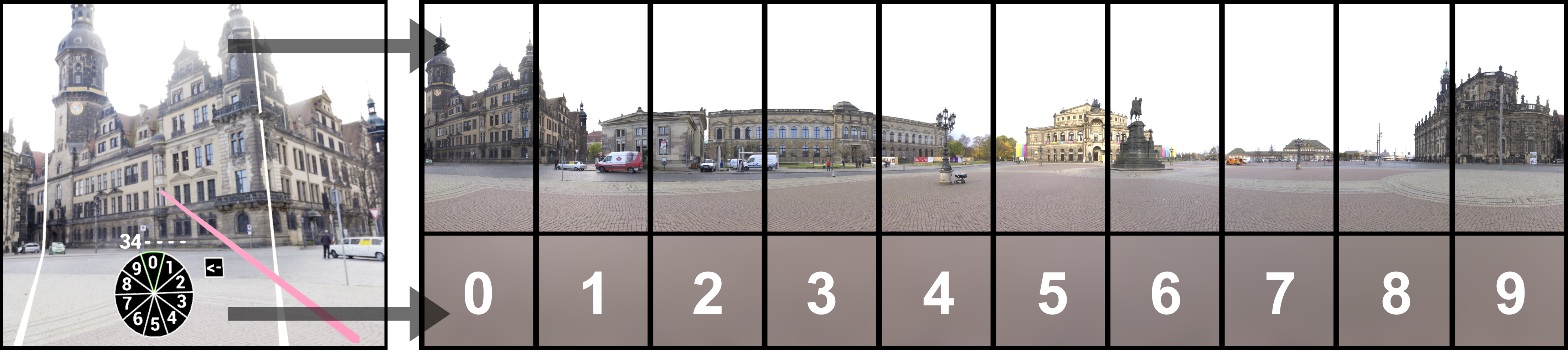}
  \caption{The Adaptive Direction-Based Authentication. The method consists of 6 virtual environments, each divided into 10 directions. Each direction is mapped to a digit from 0-9. Users can rely on the context cues for better memorability when unfamiliar with the digit sequence. They can also enter the digits directly for higher efficiency after gaining sufficient familiarity.}
  \label{fig:teaser}
\end{teaserfigure}

\received{20 February 2007}
\received[revised]{12 March 2009}
\received[accepted]{5 June 2009}

\maketitle

\section{Introduction}

% Importance of user authentication in VR
% As Virtual Reality (VR) becomes increasingly adopted in different fields, VR has the potential to play a more important role in our daily life~\cite{choi2015virtual,xie2021review,hamad2022virtual}. 
% With the envisioning of the metaverse, a lot of VR applications will be online where people are interacting, socializing, and sharing information with each other~\cite{rospigliosi2022metaverse}.
% What comes with this process is more of our personal data being collected and shared. Therefore, ensuring data security and protecting user privacy is a critical task~\cite{stephenson2022sok,de2019security,odeleye2023virtually}. 

% % Knowledge-based authentication (opportunities and challenges)
% One of the fundamental goals of privacy protection is to ensure that only the intended user can access protected resources. This is typically achieved through user authentication~\cite{idrus2013review}. 
% % p.s. can extend with brief intro to 
% Among the different types of user authentication methods, knowledge-based authentication is most widely-used due to the ease of implementation, independence of hardware, and people's familiarity with password-based systems~\cite{zviran2006identification}.
As Virtual Reality (VR) becomes increasingly adopted in different fields, ensuring data security and protecting user privacy becomes a critical task~\cite{stephenson2022sok,de2019security,odeleye2023virtually}, and many different user authentication methods have been proposed for VR~\cite{boutros2020iris,yi2016glassgesture,shen2018gaitlock,mathis2020knowledge,zhu2020blinkey,ajit2019combining}.
% p.s. can extend with brief intro to 

Knowledge-based authentication is most widely-used due to its simplicity, hardware independence, and users' familiarity~\cite{zviran2006identification}.
However, an apparent drawback is that it requires explicit memorization and entry for authentication, which introduces trade-offs between efficiency, memorability, and security~\cite{zviran2006identification}. 
For example, despite longer PINs theoretically provide stronger security by offering a larger password space, research has shown that the benefit is offset by users' tendency to choose weaker passwords~\cite{wang2017understanding,markert2020pin,munyendo2022same}.
Some authentication methods were designed to enhance memorability by utilizing picture memory, but they usually come at the cost of slower entry~\cite{george2020gazeroomlock,funk2019lookunlock}, and conversely, methods that excel at efficiency are generally less memorable, especially with strong passwords~\cite{mathis2020rubikauth}.
Moreover, the 3D nature of the spatial interaction in VR gives rise to shoulder-surfing attacks during user authentication~\cite{bovsnjak2020shoulder,christopher2022reconceptualizing} because the gestures and head movement associated with password entry can give away information about the password, and users typically lack the awareness of their surroundings when immersed in a VR headset, leaving ample opportunities for malicious attackers to exploit this vulnerability~\cite{george2019investigating,mathis2021fast,wu2023privacy}.

Very few methods have attempted to optimize efficiency, memorability, and security concurrently. The direction-based authentication (DBA)~\cite{huang2023dba, huang2024direction} combines symbol entry and context cues to offer efficient entry method and memorable cues, and utilizes randomization to counter shoulder-surfing attack. Despite promising results, the method offered symbols and contexts in a simplistic manner without considering the temporal factor in users' needs.
% The tension between efficiency, memorability, and security in VR authentication is intricate, and very few methods have been designed to optimize all of them at the same time. The direction-based authentication (DBA)~\cite{huang2023dba, huang2024direction} represents one such attempt which combines symbol entry and context cues to offer efficient entry method and memorable cues, and utilizes randomization to counter shoulder-surfing attack. The study showed promising results, but the method offered both symbols and contexts in a simplistic manner without carefully considering users' evolving needs.

We note that while both efficiency and memorability are critical for usability, they are rarely considered equally important at all times. Users are unlikely to complain about efficiency when they still struggle with memorability. Similarly, users are also unlikely still uncertain about the passwords when efficiency becomes their main concern.
Therefore, although efficiency and memorability are difficult to achieve concurrently, the conflict may be resolved by prioritizing what is most needed by the users at a given time.

Additionally, prior work on VR authentication typically adopts single- or double-session experiments in which participants complete a series of tasks in laboratories~\cite{george2017seamless,olade2020exploring,mathis2020knowledge,funk2019lookunlock}. Although such designs allow efficient data collection and better control over the procedure, they compromise ecological validity. Both the experimental tasks and settings differ significantly from real-world authentication scenarios, which raises questions of whether the results of these experiments generalize to real-world contexts~\cite{mathis2022stay}.

In this work, we tackled both research gaps identified above. Our contributions to the advancement of knowledge-based authentication in VR are as follows: 
\begin{enumerate}
    \item We introduced a novel perspective to decouple the conflict between efficiency and memorability temporally.
    \item We revised DBA to prioritize efficiency or memorability based on users' needs.
    \item We conducted a remote longitudinal study to enhance the ecological validity of evaluation by emulating real-world authentication. To our knowledge, this is the first evaluation study for VR authentication methods with such a design.
    \textcolor{red}{\item We empirically demonstrated that users do not prioritize sheer entry efficiency even under daily authentication.}
\end{enumerate}
%This work provides novel insights into the simultaneous optimization of efficiency, memorability, and security, thereby contributing to the advancement of knowledge-based authentication in VR.

% =========================================

% What have we learned from previous study/ related work
%The results of our exploratory DBA study demonstrated the advantages of contexts and symbols respectively. Context information is highly memorable, with the context-only authentication method achieving a 94\% successful recall rate after one week, even though users were not informed in advance about the recall task. Symbol-based entry method contributes to higher efficiency, supported by the findings that the two methods incorporating symbol-based entry were significantly more efficient than the context-only method. These results suggests that the context could be most helpful when the password is newly-created, but the reliance on contexts could gradually shift to symbols as users become more familiar with the password.

% In this study, we refine the original design of DBA based on findings from the previous study, and particularly to better support the distinct roles of symbols and context as outlined above. I will conduct a longitudinal study to account for changes in password familiarity over time and to gain deeper insights into the interplay between efficiency, memorability, and familiarity, with our new design. I expect the revised method to excel in efficiency, memorability, and security.

\section{Related Work}
Knowledge-based authentication in VR can be divided into symbol-based, and context-based methods depending on the type of information users need to remember~\cite{huang2023dba}.

\subsection{Symbol-Based Authentication}
% symbol-based
Symbol-based methods define passwords as a sequence of symbols, such as text or digits, and are widely used in our daily life. However, their usability in VR often relies on a small symbol set (e.g., a Numpad)~\cite{george2017seamless,olade2020exploring}, since text entry with full VR keyboard is relatively error-prone and inefficient~\cite{grubert2018text}. This reduction in symbol variety introduces security vulnerabilities to shoulder-surfing attacks~\cite{wu2023privacy,lee2023vrkeylogger,zhang2025airtypelogger} because the movements associated with the password entry become easier to exploit. 
Therefore, symbol-based authentication methods specifically designed for VR often incorporate strategies to mitigate the risk of shoulder-surfing.

One approach is to randomize the password interface~\cite{mali2015advanced,maiti2017randompad,jirjees2021roundpin}. For example, Lange et al.~\cite{lange2024vision} proposed three authentication schemes using different symbol sets with randomized layout.
Some approaches authenticate with subtler movements, such as \textit{VR Pursuit}~\cite{khamis2018vrpursuits}, which selects moving digits with continuous eye-gaze, along with other eye-gaze-based methods~\cite{rajanna2018dygazepass, abdrabou2018engage, liebers2020gaze}.
Some adopt more complicated password interfaces, such as \textit{RubikAuth}\cite{mathis2020rubikauth}, which utilizes a virtual 3D cubic PIN-pad attached to one controller.
%among other relatively more complex interface designs~\cite{yu2016exploration,yu2016usable,bu2023secure}.
%However, complex interfaces often lead to higher cognitive demands for users. 
To mitigate the usability drawback, Weiss et al.~\cite{weiss2024exploring} proposed hand redirection during PIN entry, which offsets users' actual hands by a random displacement, causing them to reach toward different locations even when choosing the same input.

 %Other examples of enhancing observation resistance with more complicated 3D password interfaces include .

Symbol-based methods are generally efficient thanks to the ease of selecting symbols on a compact PIN-pad. For example, the authentication time for RubikAuth can be under 5 seconds for a 4-digit password. However, a major shortcoming is the conflict between memorability and security as discussed earlier. RubikAuth achieves a successful recall rate of 95.24\% after a week for weak passwords, but the number plummets to 61.9\% and 42.86\% for medium and strong passwords, respectively.
%In fact, despite memorability being an important usability metric, a large number of studies on symbol-based authentication did not report it at all~\cite{noah2025pins}.

% context-based
\subsection{Context-Based Authentication}
Context-based methods authenticate using virtual environment contexts and improve memorability by leveraging the picture superiority effect, i.e., a well-established theory that people remember pictures better than words~\cite{paivio1968pictures}.
For example, in \textit{RoomLock}, the password is defined a sequence of objects in a virtual room ~\cite{george2019investigating,george2020gazeroomlock}. 
\textit{LookUnlock}~\cite{funk2019lookunlock} defines the password as a combination of virtual and real-world targets. 
\textit{3DPass} presents a virtual home environment and defines the ``password'' as a series of navigation and actions, such as entering the kitchen and turning on the lights~\cite{gurary2017leveraging}. 

While context-based methods generally achieves high memorability, the tasks associated with the authentication procedure, such as visual search and locomotion, can be inefficient. For example, the \textit{3DPass} achieved a successful recall rate of 98\% after a week, but the authentication took 21 seconds on average.

% Users generate a 3D password by performing a set of actions and navigations. Figure 2 can be considered an example 3D password. The user enters the kitchen and turns on the lights.

% dba
\subsection{Direction-Based Authentication}
To strike a balance between efficiency and memorability, Huang et al., proposed Direction-Based Authentication (DBA), which offers both symbols and contexts under a unifying framework~\cite{huang2023dba}. Passwords in DBA are defined as a sequence of directions, corresponding to both direction symbols on a compass, and environment contexts users see in the directions. While prior work demonstrated the efficiency and memorability benefits of DBA~\cite{huang2024direction}, it did not address the nuanced tradeoffs between efficiency and memorability across different usage scenarios, nor did it empirically validate the memorability advantage of DBA.

%Our consideration is to prioritize what is needed based on users' evolving familiarity with the passwords
% ---------------------------------------------------------------------
%\subsection{Environmental Reinstatement Effect}
One such nuance not explicitly considered in DBA is the temporal separation of priority between efficiency and memorability, as mentioned in the introduction.
Additionally, research has demonstrated the environmental reinstatement effect~\cite{smith2001environmental}, according to which information retrieval improves when the environmental context is consistent between encoding and retrieval~\cite{smith2013effects}.
%For example, Godden and Baddeley conducted a study on scuba divers where they learned two different lists of words, one on land and the other underwater. They found that the words they learned in a specific environment were better recalled in the same environment than in the other~\cite{godden1975context,godden1980does}.
% Discuss real world application
%This effect has practical applications in real-world settings. For instance, law enforcement often leverages environmental reinstatement by bringing eyewitnesses back to the crime scene to help facilitate memory recall.

%We find the environmental reinstatement effect to be particularly relevant to DBA where both symbols and context are offered, because the context cues can not only serve as easy-to-remember data on their own but also facilitate the memorization of their symbol counterparts by providing additional environmental context for symbol recall. Therefore, the context cues may foster the transition to using symbols for higher efficiency in the long run while providing a memorable fallback.
%This is in line with our envisioning that the conflict between efficiency and memorability can be solved by prioritizing what is most needed by the users depending on how familiar they are with the password.

Based on these insights, we noted that the context cues can not only serve as easy-to-remember data on their own but may also facilitate the memorization of their symbol counterparts by providing additional environmental context for symbol recall. 
%Therefore, the context cues may foster the transition to using symbols for higher efficiency in the long run while providing a memorable fallback. 
This is in line with our envisioning that the conflict between efficiency and memorability can be resolved temporally by prioritizing what is most needed by the users at any given stage.
We revised DBA and conducted a longitudinal study with a remote design emphasizing the ecological validity of the experiment for evaluation.

%Therefore, it might be beneficial to rely on context for password creation to fully utilize its memorability. As users enter the password, they naturally develop context-dependent memory of the symbol representation, as the symbols are consistently displayed alongside the context. Once users become sufficiently familiar with the symbol representation, they can transition to entering the symbols directly, enhancing efficiency.

\section{The Adaptive Direction-Based Authentication}
We present Adaptive Direction-Based Authentication, or ADBA, which is built upon DBA with revisions aimed at adapting to users' priorities.
ADBA consists of a sequence of 6 virtual environments, each divided into 10 directions corresponding to the digits 0 through 9. Users create their password by selecting one direction in each environment. See Figure~\ref{fig:teaser} for visual illustration.
%resulting in a password space of $10^{6}$, aligning with the widely-used ``4-digit PIN" scheme. For scenarios requiring a larger password space, the method can be easily extended to include additional virtual environments without compromising its general applicability.

\paragraph{Password Creation}
In contrast to the original DBA where users can choose either symbol or context, ADBA requires users to create passwords based on context by displaying only the virtual environment but not the digits corresponding to each direction. 
% Interaction
Users point their controller at a direction to select it. A glossy frame will highlight the part of the context corresponding to the selected direction. 
They can then press trigger button on the controller to confirm their selected direction as part of their password.

% Design Justification
This enforced context-based password creation offers both memorability and security benefits. The context is inherently easier to remember due to the picture superiority effect, which is more important than a quick yet potentially less memorable password setup. 
Additionally, since the users do not know the digit correspondence during password setup, they cannot choose weak passwords like 123456. Instead, we generate random digit sequences as the digit representations to enhance security. 
%Even though the random sequence could be difficult to remember,  

%users are unable to choose weak, predictable passwords, such as 123456, since the underlying digits are hidden during password creation. Consequently, the numeric representation of the chosen password becomes effectively random, achieving the maximum security theoretically.

\paragraph{Password Entry}
% Interface
During authentication, users are presented with both the virtual environment and a virtual radial PIN-pad that floats in front of them. The PIN-pad provides an alternative ``symbol-based'' entry method, and rotates as the users physically turn, similar to a compass, ensuring that the digit corresponding to the forward direction is always in front. This emphasizes the correspondence between the directions and the digits, and offers additional security against shoulder-surfing attack because the digits are not in fixed locations as in standard PIN-pads.
Above the PIN-pad, a sequence of digits corresponding to the directions entered in the previous virtual environments is displayed.
%, allowing users to track their input and ensure accuracy throughout the process.

% Interaction
Users can select either digits or contexts depending on whether they point at the PIN-pad or the environment. In either case, both the digit and the context corresponding to the selected direction will be highlighted. Users press the trigger button to confirm selection.
A backspace button is also provided for correcting mis-entries.

% Design Justification
By continuing to provide context-based entry, ADBA supports users continuing to prioritize memorability over efficiency.
By providing additional symbol-based entry, ADBA enables a more efficient alternative for users who has already remembered the digit representations and choose to prioritize efficiency.
Additionally, the presence of the context may also cue the recall of the corresponding digit, which facilitates memorization.
Overall, ADBA is a context-only method during password setup, and either remains context-only, transitions into symbol-only, or exists as a hybrid method during authentication depending on users' varying needs.

%Over time, depending on users' preferences or memory capacity, they may choose to either keep using context-based entry or leverage the association between the context and its corresponding digit representation to switch to using digits for faster entry. At this stage, the authentication either remains context-based or transitions into a symbol-based approach.

%It is important to clarify that while ADBA offers digit-based entry as an option, it is not intended to push all users toward adopting digits over contextual cues. \textcolor{red}{Users should be enabled to use whatever they prefer rather than required to prioritize efficiency.}

%----------------------------------------------------------------------

\section{Methods}

\begin{figure}[t]
  \centering
  \includegraphics[width=0.95\columnwidth, trim = 0px 30px 0px 0px, clip]{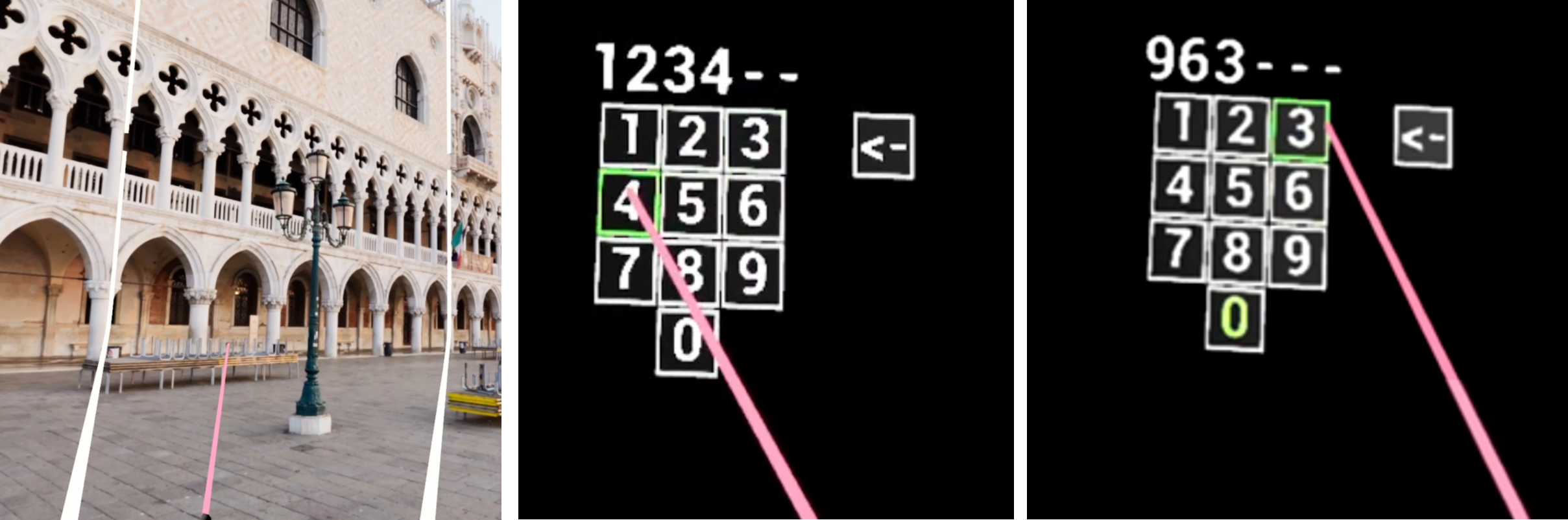}
  \caption[]{ADBA (left), 6-digit-PIN (middle), and random-PIN (right) during password setup. Note that the radial PIN-pad is hidden during password setup with ADBA. The two PIN-based methods share the same interface, but for random-PIN the next digit to choose is randomly-determined and highlighted in green (0 in the screenshot).
    }
  \label{fig:conditions}
\end{figure}

\subsection{Study Design}
% General Info (remote + longitudinal)
We conducted a between-subject remote longitudinal study to evaluate the efficiency, memorability, and security of ADBA. 
The between-subject design reduces the interference between conditions, and the remote and longitudinal design allows participants to authenticate in their own environments, at their own pace, with their personal setups, and at realistic frequencies. These together contribute to a scenario that closely resembles the real-world authentication compared to within-subject or lab-based experiments.

% Formal Study
The study consists of three conditions: 1) ADBA, 2) user-selected 6-digit-PIN and 3) randomly assigned 6-digit-PIN. See Figure~\ref{fig:conditions} for visual illustrations of the three conditions.
We chose to compare ADBA against PIN-based approaches because they are widely adopted and remain more efficient than the recent knowledge-based schemes.
They are also structurally comparable given the identical theoretical password space ($10^{6}$).
While six random digits are impractical for adoption due to memorization difficulty, they represent an ideal security scenario for 6-digit-PINs, identical to the digit representation of ADBA except without context information.

%whereas comparing ADBA against 6-Digit-PIN is a more direct comparison with a widely-adopted authentication method.

\subsection{Materials}
The study software was developed using Unreal Engine 5.3, and deployed as a PC-based VR application compatible with both steam VR and Meta Quest Link.
%The authentication method is configured using a text file, provided to the participants during the first session.
The app would automatically upload data to the server every time a participant completed a password entry on their end.
The virtual environments were represented by 360\textdegree\space high dynamic range images (HDRIs) obtained from HDRI haven~\cite{hrdihaven}.

\subsection{Procedure}

% Screening and recruitment
Participants first completed the screening, verified software compatibility with their setup, and provided informed consent.

The study had 4 stages that spanned across 2-3 weeks:
\begin{itemize}
    \item \textbf{Password Setup}: Participants scheduled a Zoom meeting, during which the experimenter explained the instructions and the participants set up their password.
    \item \textbf{Low Frequency Entry}: Participants entered the password every 6-8 days for two times.
    \item \textbf{High Frequency Entry}: Participants entered the password every 1-2 days for four times.
    \item \textbf{Exit Survey}: Participants filled out the usability and demographics surveys.
\end{itemize}

% Password creation
For password setup, participants assigned to the ADBA condition were instructed to check out all directions before selecting a direction. They were also advised to focus on less noticeable details. The underlying digit representations were six randomly generated, non-sequential, non-repeated digits \textcolor{red}{(e.g., no "123456" or "123246")}.
For consistency, participants assigned to the user-selected 6-digit PIN condition were required to choose distinct, non-sequential 6-digit sequence. They were also instructed not to use their real-world PINs for their own privacy. 
The randomly-generated PINs follow the same rule. Participants assigned to this condition selected the pre-determined digits highlighted in green during setup.
\textcolor{red}{To confirm memorization upon creation}, participants must correctly enter their passwords three times to proceed.
Participants were also be explicitly instructed not to write down their passwords. To minimized the incentive for participants to cheat, they were informed that, while they should try their best to remember the passwords, forgetting had no negative impact on their compensation and they would simply be reminded of the password if that happened. 
%However, they would lose the compensation if they cheated.

% Password entry
Over the subsequent weeks, participants performed authentication tasks at their convenience following the required frequencies. No pre-schedule was needed as data collection happened automatically.
We considered both a low and a high frequency because frequency may affect users' preferences or priorities between efficiency and memorability.
Participants entered the recovery mode after 3 failed authentication attempts. In the PIN-based conditions, the correct digits were highlighted in green. In the ADBA condition, the digits were hidden and the correct directions were indicated by a green frame.
%Only correct entries were accepted during the recovery, and the non-highlighted options cannot be entered. 
We decided to remind instead of reset passwords to ensure all participants remembering a single password throughout, avoiding confounding factors for memorability.
%them of their password instead of asking them to create a new one because having to create and remember multiple passwords during different stages of the study can introduce confounding factors for memorability.

\subsection{Participants}
66 participants (18F, 46M, 2NB, aged between 18 and 74) were recruited and completed the study, among which 33 were recruited through Prolific, and 33 from the university community.
The sample size was calculated using G*Power with a large effect size of 0.4 and to achieve a power of 0.8 for a fixed-effects, omnibus one-way ANOVA.
Participants were required to have access to a PC-VR setup for the duration of the study. 
%They received \$3 after completing the initial zoom session, \$1 after completing each of the 6 password entry tasks, and a \$11 completion bonus, totaling to \$20 if they completed all parts of the study.
They received \$20 if they completed all parts of the study, prorated if they dropped out early.
The study was approved by the university's institutional review board (IRB).
%They will be compensated at the minimal rate allowed by Prolific, but will be given a bonus that will make the average compensation $\$15$ per hour if they complete the entire study.

Due to the longitudinal nature of the study, some participants ran into unforeseeable scheduling conflicts. We decided to accept submissions that were less than 24 hours early or late. Only 6 out of the 462 task submissions fell into this category.
%from participants completing the study, 5 were late for less than 24 hours and 1 was early for less than 24 hours.
%In additional to the 66 participants, some participants provided valid data for some sessions but did not make it to the end of the study. Their valid data was included in data analysis where its inclusion does not violate the assumptions of the statistical tests. More details on the data analysis are provided in section \ref{sec:results}.

\subsection{Measures}
\paragraph{Entry Efficiency}
We recorded the time it took for each password entry, from the time participant pressed the start button to load the authentication interface to the time the last entry was completed. %Unlike many prior works that excluded the time before the first entry is made~\cite{mathis2021fast,george2019investigating,george2017seamless}, we decided to include it because the search time for context-based authentication methods is nontrivial and excluding it can lead to significant under-estimation.

%\textcolor{red}{consider adding authentication time on top of entry time}

\paragraph{Memorability}
We evaluated memorability using both the \textit{recall rate}
%, i.e., number of successful authentication (in 3 attempts) divided by the total number of authentication and 
and the \textit{number of entries per authentication}.
Since users were allowed to correct their input, submitting incorrect passwords is more likely due to memorability issues rather than input errors.

\paragraph{Usability and Task Load}
We administered the System Usability Scale (SUS)~\cite{brooke1996sus} and the raw NASA TLX~\cite{hart2006nasa} surveys.
We also included a custom usability survey in which we asked participants to indicate %the extent to which they agree with the two statements \textit{``I was able to authenticate efficiently.''} and \textit{``I was able to remember the password well.''} 
their perceived efficiency and memorability in 7-point likert scale. 
For participants assigned to the ADBA condition, we asked if they tried to remember the digits associated with each environment. If they responded yes, we asked them to indicate 
%their agreement to the statement \textit{``It required significant effort to remember the digits.''} 
the perceived effort required to remember the digits in 7-point likert scale.
Lastly we asked participants for open-ended feedback.

\paragraph{Password Homogeneity}
We focused on password homogeneity as the primary security indicator, as this was identified as a key potential vulnerability in DBA~\cite{huang2024direction}. Specifically, we computed the Shannon entropy of the directions selected by users within each environment, where lower values suggest more homogeneous choices.
%
%Higher entropy values indicate greater diversity in password choices, while lower values suggest a tendency to favor certain directions.
%and to a greater extent than in the two PIN-based methods, which could enable attackers to make more informed guesses.
%To address the limited password space noted in the original method~\cite{huang2024direction}, we expanded the design by increasing the number of divisions per environment from 8 to 10 and the number of environments from 4 to 6. Additionally, in ADBA, the digit representation of each password is randomly generated, thereby mitigating the risk of users creating weak or predictable digit sequences.
%
While shoulder-surfing resistance is also an important security aspect, we omitted it due to practicality and potential privacy concerns in the remote setup. Additionally, the robustness of similar interfaces has already been demonstrated in~\cite{huang2024direction,huang2026secure}.

\subsection{Hypotheses}
Our hypotheses are as follows:
\begin{itemize}
    \item \textbf{H1:} Participants would remember the password better with ADBA compared to both PIN-based methods.
    \item \textbf{H2:} There would exist an interaction effect where ADBA would be the slowest from the start, but the difference would diminish for the last entry.
    \item \textbf{H3:} Participants would report better usability with ADBA compared to the PIN-based methods.
    \item \textbf{H4:} Participants would report lower task load with ADBA compared to the PIN-based methods.
\end{itemize}

% These hypotheses are grounded in the design features of ADBA. Since the ADBA provides additional context cues to facilitate memorization, we expected it to yield better memorability than the PIN-based methods, where users rely solely on digits (H1).
% %
% Since searching for directions is inherently slower than entering digits, we expected participants to initially experience longer entry times with ADBA. However, as they become familiar with the digit sequences, particularly in the high-frequency stage, we anticipated a shift toward more efficient input, ultimately resulting in comparable entry times to the two PIN-based methods (H2). 
% %
% Furthermore, we anticipated that the flexibility provided by ADBA which allows users to choose between contextual and symbolic input would result in higher perceived usability (H3) and lower reported task load (H4), as participants could authenticate in the manner most comfortable to them.
% %
% We did not formally define hypotheses related to security. The analysis of password homogeneity was exploratory, and the other security properties of ADBA have either been established in prior studies or do not require human-subject evaluation.
We hypothesized that ADBA would be the most memorable due to picture superiority effect (H1).
We expected participants to initially experience slower entries with ADBA when searching for context, but the difference may diminish if they prioritized efficiency and shifted to digit-based entry during high frequency entries (H2). 
We anticipated that ADBA would result in higher perceived usability (H3) and lower reported task load (H4), as participants could authenticate in a most comfortable manner.

\subsection{Data Analysis}
We used Bayesian generalized linear mixed model (Bayesian GLMM) analysis for multi-observations, and Bayesian generalized linear models (Bayesian GLMs) analyses for single-observation. \textcolor{red}{The analyses were performed using the 'brms' package (2.20.1) in R (4.3.1.).}

We adopted GLMMs because they capture the hierarchical structure of our data: each participant completed varying number of entries across multiple sessions. A Bayesian approach was preferred over frequentist methods because it performs more robustly with small-to-medium sample sizes (22 per condition) and yields a more informative continuous estimates of posterior probabilities.

\section{Results}
\label{sec:results}

We report descriptive statistics as mean (SD) when Shapiro–Wilk tests do not indicate departures from normality; otherwise we report median (IQR). 
\textcolor{red}{The reported results were achieved using brm default bayesian prior (flat priors for population-level regression coefficients and Student-t priors for intercepts and group-level standard deviations). For sensitivity test, we used weakly informed priors for coefficients and the resulting posterior probabilities were generally similar to those obtained under the default priors, and substantive conclusions remained unchanged.
Models were estimated using four chains with 4,000 iterations per chain, including 1,000 warm-up iterations, using the NUTS sampler.} All regression models achieved good convergence ($\hat{R} = 1$). Effects were considered reliable if their 95\% highest posterior density (HPD) intervals did not include zero. 
We report all effects with posterior probability $>$ 85\%, as Bayesian results do not rely on a binary cutoff.

\subsection{Memorability}
\label{sec:memorability}

\begin{figure}[t]
  \centering
  \includegraphics[width=0.9\columnwidth, trim = 0px 10px 0px 7px, clip]{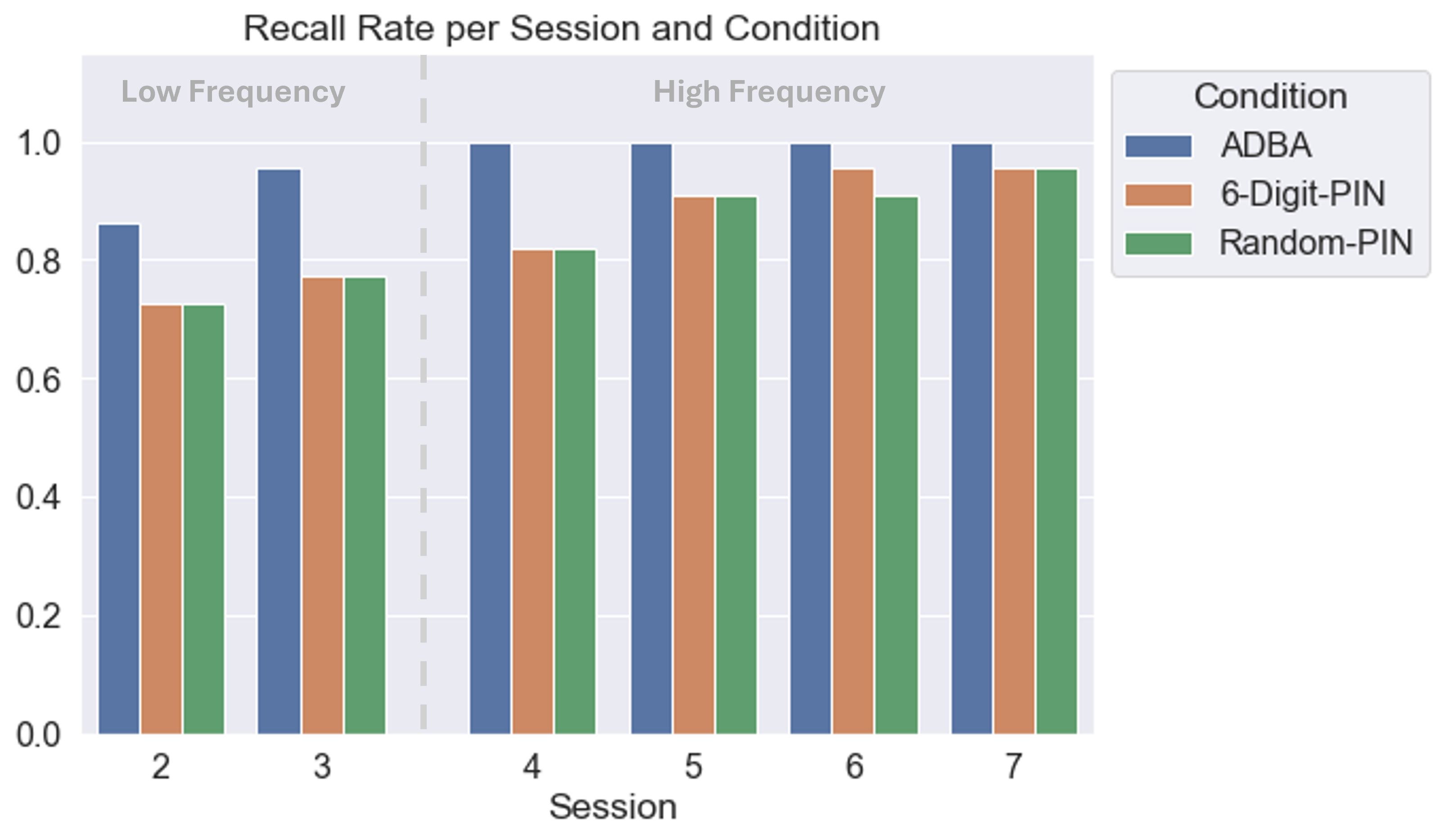}
  \caption[]{Recall rate per session. ADBA yielded higher recall rate consistently. Bayesian GLMM indicated P(ADBA $>$ 6-Digit-PIN) = 0.92 and P(ADBA $>$ Random-PIN) = 0.95 during the low-frequency entry stage.
    }
  \label{fig:recall}
\end{figure}

We present the memorability results across all sessions to illustrate the overall trend. However, the statistical analysis focuses on the low-frequency entry stage because memorability is most critical when passwords are newly created and infrequently used. 
%Including high-frequency entry data would add noise, as most participants were able to recall their passwords during that stage regardless of condition.

\paragraph{Password Recall}
Figure \ref{fig:recall} shows the recall rate per session.
Recall rate for ADBA was the highest (Session 2: 86\%; Session 3: 95\%), while the 6-Digit-PIN and Random-PIN conditions showed identical performance (Session 2: 73\%; Session 3: 77\%).

A Bayesian GLMM with a Bernoulli likelihood and logit link was fitted to participants’ binary recall outcomes, with \textit{Condition} and \textit{Session} as fixed effects, and random intercepts for participant ID.
The posterior median estimates for condition effects, relative to ADBA, were -2.81, 95\% HPD [-7.44, 0.71] for 6-Digit-PIN and -3.14, [-7.87, 0.18] for Random-PIN, both on the log-odds scale.

Pairwise contrasts indicated that ADBA yielded higher recall rates than both 6-Digit-PIN (median log-odds difference = 2.63, [-0.96, 6.90]) and Random-PIN (2.88, [-0.25, 7.73]). Although the 95\% HPD intervals included zero, the posterior probabilities that ADBA outperformed the other two conditions were:
P(ADBA $>$ 6-Digit-PIN) = 0.92 and P(ADBA $>$ Random-PIN) = 0.95. These results support H1 that participants were able to recall better in ADBA.

\begin{figure}[t]
  \centering
  \includegraphics[width=0.9\columnwidth]{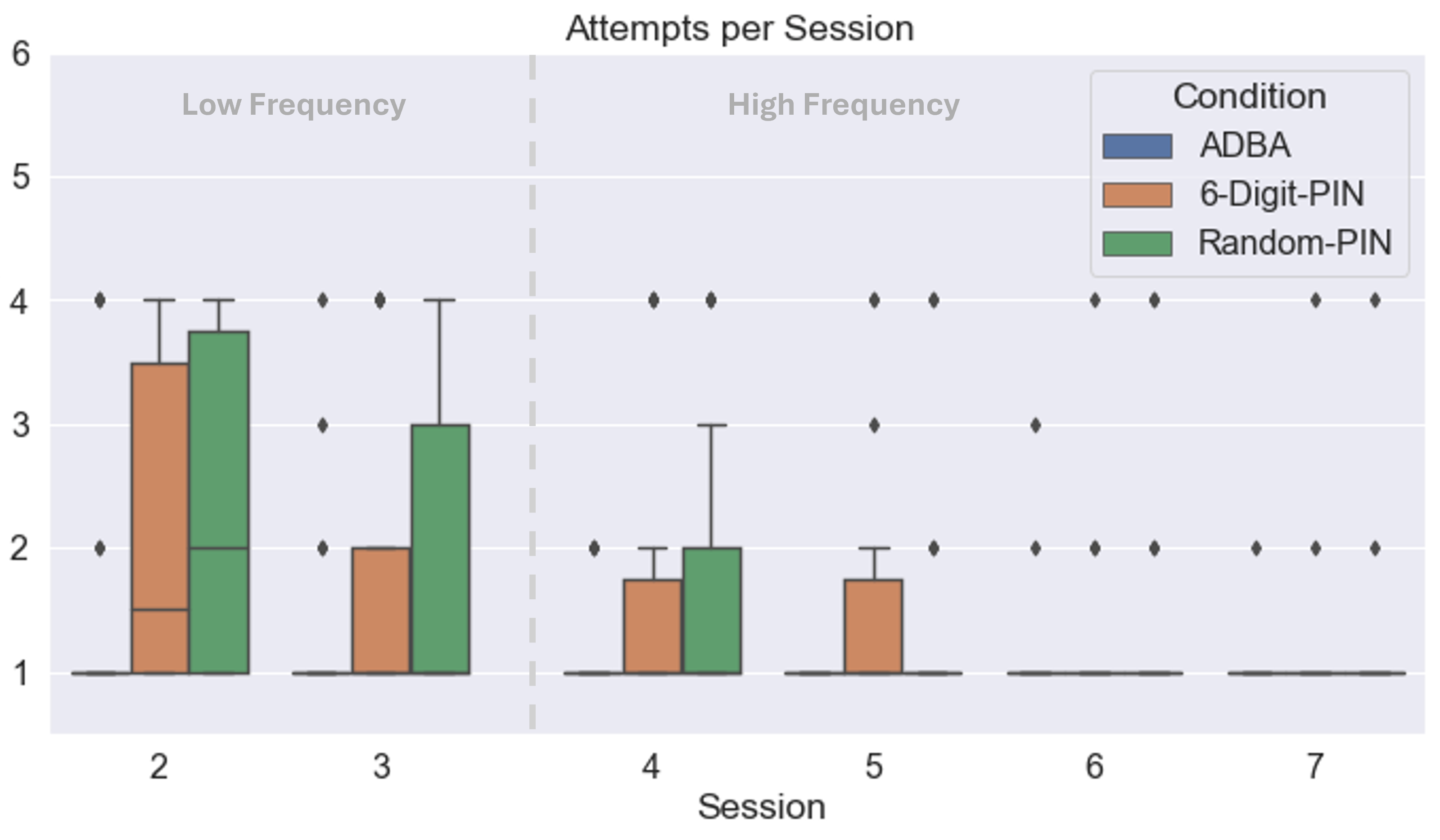}
  \caption[]{Number of attempts per session. Bayesian GLMM indicated P(ADBA $<$ 6-Digit-PIN) = 0.92 and P(ADBA $<$ Random-PIN) = 0.97 during the low-frequency entry stage.
    }
  \label{fig:attempt}
\end{figure}

\paragraph{Entry per Session}
Figure \ref{fig:attempt} shows the number of entries per session. 
ADBA required the fewest entries ($Mdn = 1, IQR = 0$ in both session 2 and 3), followed by 6-digit-PIN (Session 2: $Mdn = 1.5, IQR = 2.5$; Session 3: $Mdn = 1, IQR = 1$), and Random-PIN (Session 2: $Mdn = 2, IQR = 2.75$; Session 3: $Mdn = 1, IQR = 2$).
%Note that the minimum number of entries per session was 1 and the maximum was 4 (three failed attempts plus one password recovery).

A Bayesian Poisson mixed-effects model was fit to participants’ entry counts, with \textit{Condition} and \textit{Session} as fixed effects and random intercepts for participant ID.
The posterior median estimates for condition effects, relative to ADBA, were 0.30, [-0.06, 0.67] for 6-Digit-PIN and 0.39, [0.05, 0.75] for Random-PIN.

Posterior pairwise comparisons revealed that ADBA required fewer entries than Random-PIN (Mdn difference = -0.39, [-0.75, -0.05]) and 6-Digit-PIN (-0.30, [-0.67, 0.07]).
The posterior probability that DBA required fewer entries were P(ADBA $<$ 6-Digit-PIN) = 0.92 and P(ADBA $<$ Random-PIN) = 0.97. This result also supports H1 that ADBA required fewer attempts to authenticate successfully.

\subsection{Efficiency}
\label{sec:res-efficiency}
%\textcolor{red}{consider adding authentication time on top of entry time}

\begin{figure}[t]
  \centering
  \includegraphics[width=0.9\columnwidth]{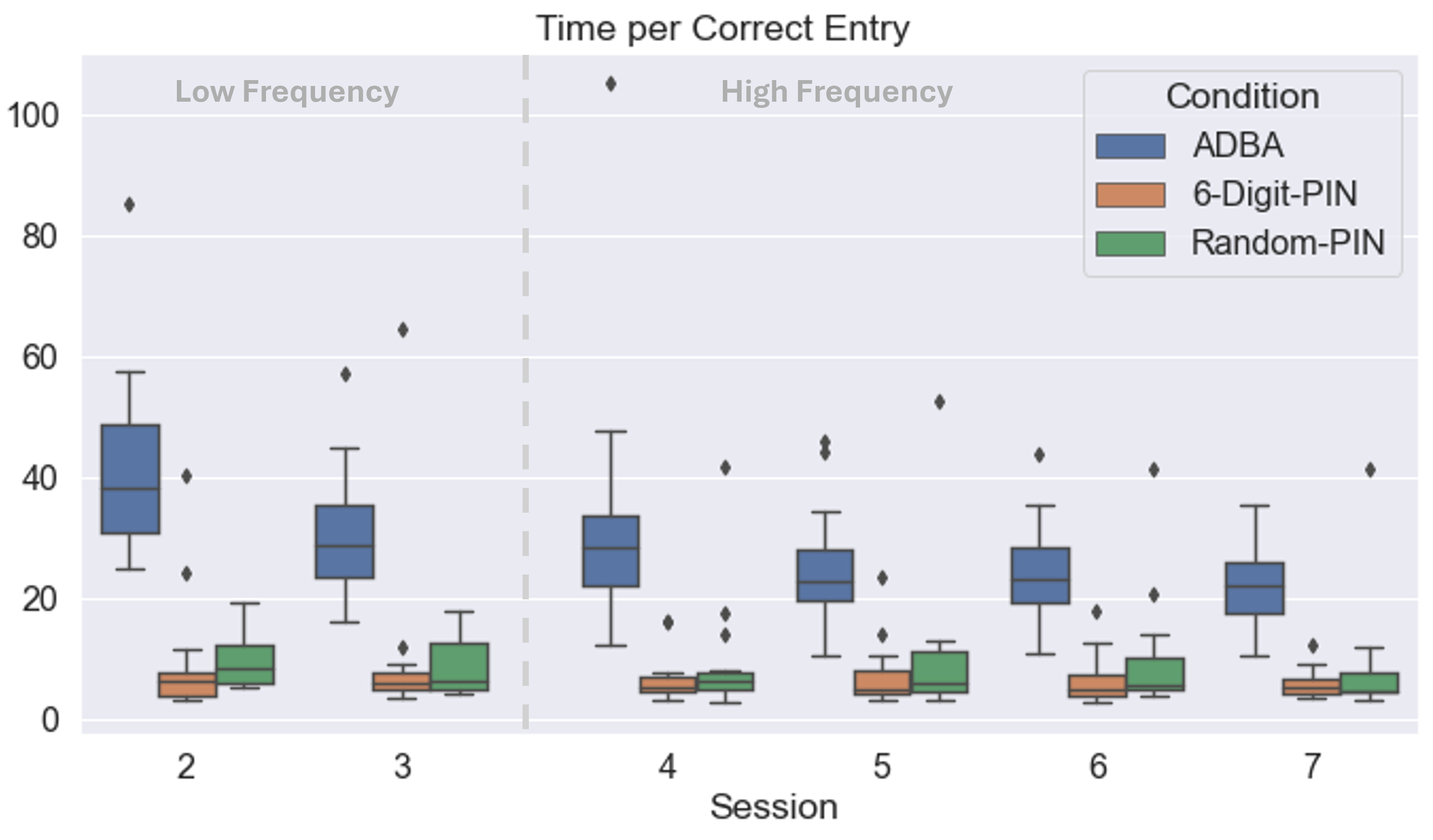}
  \caption[]{Correct entry time per session. Bayesian GLMM indicated main effects of Condition (ADBA was slower), Session (Entry time decreased over sessions), and the interaction (ADBA entry time decreased faster over session).
    }
  \label{fig:entry}
\end{figure}

Figure \ref{fig:entry} shows the entry time per session. We included the correct entries only, excluding entries during password setup and recovery.

Across all six entry sessions, the ADBA condition yielded the slowest entries ($Mdn = 25.2, IQR = 12.9$ secs), whereas the 6-Digit-PIN ($Mdn = 5.5, IQR = 3.4$ secs) and Random-PIN ($Mdn = 5.9, IQR = 3.6$ secs) conditions were substantially faster.

A Bayesian linear mixed-effects model was fit to entry times, with \textit{Condition}, \textit{Session}, and their interaction as fixed effects, and random intercepts for participant ID. %The model showed good convergence (all $\hat{R} = 1.00$; effective sample sizes > 1700).
The posterior median estimates indicated a strong main effect of condition: relative to ADBA, both the 6-Digit-PIN (-21.52, [-24.82, -18.28]) and Random-PIN (-19.95, [-23.30, -16.57]) showed substantially shorter entry times.
Entry time also decreased across sessions (-14.34, [-18.00, -10.59]), with notable interaction effects: entry times for ADBA decreased more steeply than for the other two methods (6-Digit-PIN × Session: 11.17, [5.67, 16.54]; Random-PIN × Session: 13.12, [7.59, 18.63]).

Posterior contrasts confirmed that ADBA was consistently slower than both PIN-based methods in all sessions (median difference range: 14.4-32.1 seconds, all 95\% HPD intervals excluding zero). However, ADBA demonstrated a faster improvement rate.

Interestingly, only 2 participants in the ADBA condition indicated that they attempted to remember the digits. They further indicated ``disagree'' and ``somewhat disagree'' to the statement \textit{``It was difficult to remember the digits''}. %answered ``yes'' to the question \textit{``Did you try to remember the digits associated with each environment?''}. These participants further responded ``disagree'' and ``somewhat disagree'', respectively, to the statement \textit{``It was difficult to remember the digits''}. 
Notably, neither switched to using digits as their primary entry method.

These findings partially support H2, as a significant interaction effect between condition and session was observed. However, this improvement was not sufficient to close the performance gap in entry speed by the end of the study, because no participant in the ADBA condition switched to the digit-based entry.

\subsection{Usability}

\paragraph{System Usability Scale} The SUS scores were similar across all conditions: 6-Digit-PIN ($Mdn = 87.5, IQR = 10.0$); ADBA ($Mdn = 83.8, IQR = 18.1$); Random-PIN ($Mdn = 83.8, IQR = 15.0$).

A Bayesian linear model revealed no credible differences across conditions: ADBA - 6-Digit-PIN = -1.79 [-9.42, 5.30]; ADBA - Random-PIN = 0.52 [-6.93, 7.94]; 6-Digit-PIN - Random-PIN = 2.42 [-5.28, 9.48]. These results do not support H3.

\paragraph{Perceived Efficiency and Memorability}

\begin{figure*}[t]
  \centering
  \includegraphics[width=0.9\textwidth, trim = 0px 8px 0px 6px, clip]{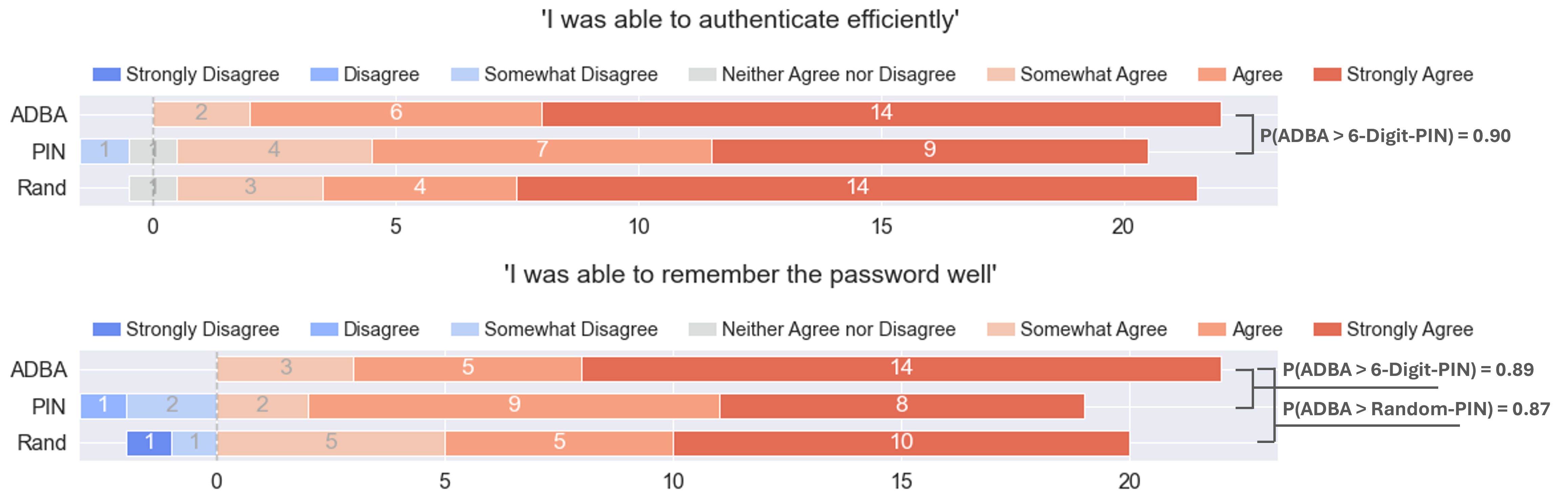}
  \caption[]{Likert-type responses to \textit{``I was able to authentication efficiently''} and \textit{``I was able to remember the password well''}. Bayesian GLMM indicated better subjective efficiency of ADBA than 6-digit-PIN, and better subjective memorability of ADBA than both PIN-based conditions.
    }
  \label{fig:subjective}
\end{figure*}

Figure \ref{fig:subjective} summarizes participants' perceived efficiency and memorability.
Bayesian ordinal logistic regression models revealed higher perceived efficiency for ADBA than for 6-digit-PIN (Mdn difference = 1.02, [-0.08, 2.21], P(6-digit-PIN $<$ ADBA) = 0.90), contrary to its slower entry time.

ADBA was perceived as more memorable than 6-digit-PIN (Mdn difference = 1.00,  [-0.12, 2.17], P(ADBA $>$ 6-digit-PIN)  = 0.89), and Random-PIN (Mdn difference = 0.905,  [-0.312, 2.03], P(ADBA $>$ Random-PIN) = 0.87), consistent with objective memorability.

These findings support H3, although the 95\% credible intervals for all contrasts included zero, indicating some uncertainty.

\subsection{Task Load}

\begin{figure*}[t]
  \centering
  \includegraphics[width=0.9\textwidth, trim = 0px 2px 0px 2px, clip]{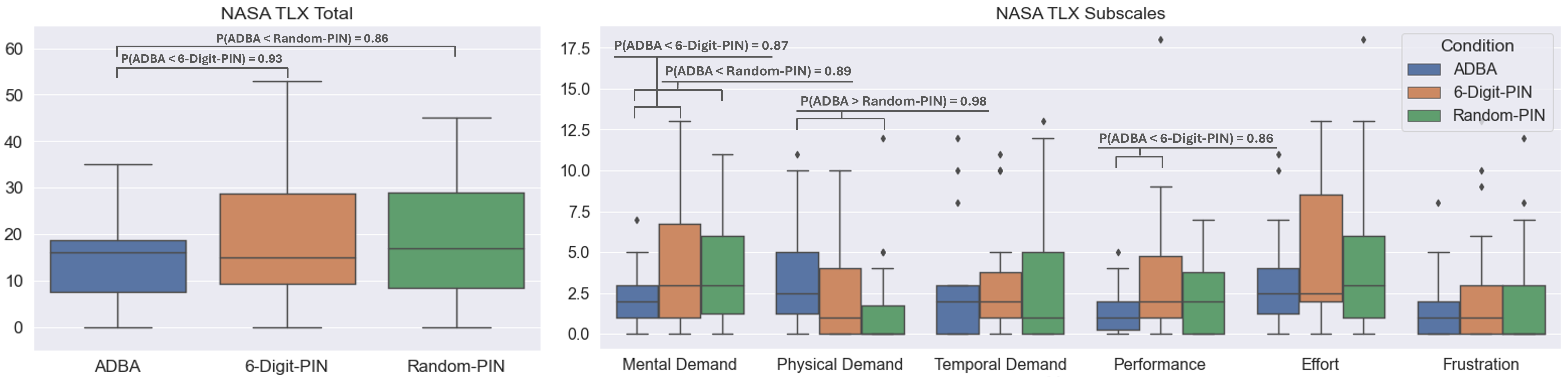}
  \caption[]{NASA TLX Scores and Subscale Scores. Bayesian GLMM indicated lower perceived task load with ADBA, particularly regarding lower mental demand than both 6-digit-PIN and random-PIN, and better performance than 6-digit-PIN. However, the reported physical demand for ADBA was higher than Random-PIN.
    }
  \label{fig:tlx}
\end{figure*}

Figure \ref{fig:tlx} shows the NASA TLX result and a breakdown of the subscales. Participants reported the lowest perceived task load for ADBA ($M = 14.5 , SD = 9.5$), followed by the Random-PIN ($M = 18.5, SD = 12.0$), and the 6-digit-PIN ($M = 20.0, SD = 15.6$).

A Bayesian linear model indicated lower perceived task load for ADBA than for the PIN-based methods. The posterior differences were ADBA - 6-Digit-PIN = -5.54, [-13.3, 1.92], and ADBA - Random-PIN = -4.05, [-11.7, 3.41]. Posterior probabilities suggested moderate evidence of lower workload for ADBA: P(ADBA $<$ 6-Digit-PIN) = 0.93 and P(ADBA $<$ Random-PIN) = 0.86. These result supports H4.

Bayesian ordinal logistic regression models for TLX subscales indicated strong evidence that ADBA was associated with higher physical demand than Random-PIN (P(ADBA $>$ Random-PIN)  = 0.98), and weak evidence of advantages for ADBA in mental demand
(P(ADBA $<$ 6-Digit-PIN) = 0.87;
P(ADBA $<$ Random-PIN)  = 0.89) 
and performance:
P(ADBA $<$ 6-Digit-PIN) = 0.86.

\subsection{Security}
\label{sec:res-security}

\paragraph{Password Homogeneity}

\begin{table}[h]
\centering   
\begin{tabular}{ |c|c|c|c|c|c|c| } 
 \hline
 Environment & 1 & 2 & 3 & 4 & 5 & 6  \\
 \hline
 Entropy (bits) & 2.56 & 2.18 & 2.77 & 2.36 & 2.61 & 2.70  \\ 
 \hline
\end{tabular}
\caption{Shannon Entropy per Environment.}
\label{table:entropy}
\end{table}

\begin{figure}[t]
  \centering
  \includegraphics[width=0.8\columnwidth, trim = 0px 13px 0px 7px, clip]{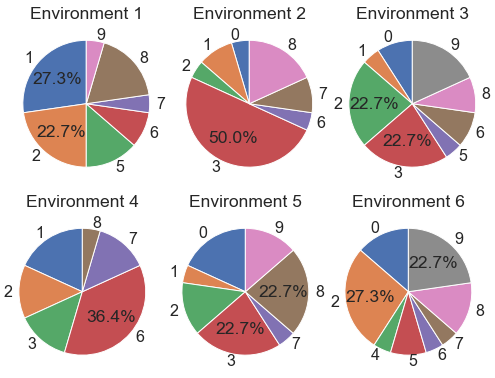}
  \caption[]{Summary of chosen directions. The digits represent the direction indices. The percentages indicate the proportion of participants choosing the corresponding direction. Environment 2 and 4 demonstrated noticeable homogeneity.
    }
  \label{fig:pw}
\end{figure}

Figure \ref{fig:pw} visualizes the selected directions in each environment in ADBA.
The Shannon Entropy for each environment is summarized in table \ref{table:entropy}.
For reference, the maximum possible entropy for a 10-digit system is $\log_2 10 \approx 3.322$ bits. The entropy for each digit in the 6-digit-PIN ranged from 2.81 to 3.14 bits, and for Random-PIN, from 2.99 to 3.17 bits.
Both the figure and the entropy values indicate that Environments 2 and 4 were associated with relatively low entropy. %, with directions 3 and 6 being popular choices, respectively. 
Environments 3 and 6 exhibited higher entropy, although still lower than the entropy observed in the 6-Digit-PIN condition. These results suggest that participants’ choices of directions within each of the six environments were less diverse than choices of digits in the 6-digit PINs.

% Entropy per Digit
% 
% DBA
% Digit 1    2.556054
% Digit 2    2.184261
% Digit 3    2.767646
% Digit 4    2.356492
% Digit 5    2.608111
% Digit 6    2.703559
% 
% 6-Digit-PIN
% Digit 1    2.806499
% Digit 2    2.901947
% Digit 3    2.992856
% Digit 4    2.828781
% Digit 5    3.044913
% Digit 6    3.135822
% 
% Random-PIN
% Digit 1    3.118078
% Digit 2    2.992856
% Digit 3    3.174674
% Digit 4    3.079226
% Digit 5    2.988317
% Digit 6    3.118078
% 

\paragraph{Vulnerability of 6-digit-PINs}

% Unlike ABDA where each environment is unique, 6-digit-PIN is reusing the same set of digits (0-9) at all 6 positions, so preference of a certain digit at a specific position in isolation is unlikely, and the risk of password homogeneity is minimal.

Despite instructions to choose secure passwords, participants still showed a preference for weak passwords. A large portion of the passwords exhibits recognizable patterns, such as “012378”, “123475”, “782345”, “547632”, etc.
Additionally, several participants asked if they could use memorable information such as birth dates or phone numbers as passwords. While this was allowed per study design and not recognizable to the experimenter, they may introduce additional vulnerabilities.

%This tendency was most commonly observed in sequences of adjacent digits, such as “012378”, “123475”, and “782345”, as well as in patterns with groups of two adjacent digits or digits with regular intervals, such as “547632”, “784512”, “467821”, “781254”, and “246813”. 
%The full list of user-selected and assigned passwords is provided in the supplementary materials.

%Another source of vulnerability arose from participants using personally memorable information, such as birth dates, phone numbers, or other familiar sequences, as their passwords. Several participants explicitly asked whether they could use such information, and this was allowed per study design, as it reflects common real-world practices and there is no practical way to prevent users from selecting these sequences. Although these passwords may appear random to the experimenter, they remain vulnerable to attackers able to obtain those personal information.

\subsection{Open-Ended Feedback}
\label{sec:qualitative}
%We conducted a qualitative analysis of the open-ended feedback provided by participants. The first author led the identification and synthesis of the main themes, which were subsequently reviewed and refined by the second and third co-authors. Any disagreements were resolved through discussion.

%Due to the between-subjects design, participants were unaware of the other conditions and thus could not directly compare or contrast the three methods. 
Among the feedback provided by the 66 participants, 26 addressed memorability, 10 were related to interaction and visualization, 6 discussed technical difficulties, and 4 mentioned security. Notably, no participants commented on efficiency.

Most participant responses focused on the their memorization strategies. Interestingly, similar approaches were reported for the two PIN-based conditions. 3 participants in 6-digit-PIN condition and 5 in random-PIN condition mentioned using spatial relationships to facilitate memorization rather than relying solely on the digits. Examples include: \textit{``My strategy was to pick a digit combination with geometric symmetry''}, \textit{``The PIN pad was definitely the reason I was able to remember the code. I used the visualization of the button placements to memorize the code''}.
Another commonly reported memorization strategy was repetition. Several participants noted that they mentally repeated the digits regularly. For instance: \textit{``I tried to think about the password even when I didn't need to enter it''}, \textit{``my brain naturally kept repeating the number at the end of the day''}, \textit{``the first weeks I would occasionally remember the study (when unlocking my phone for example) and repeat to myself the password''}. %One participant from the 6-digit-PIN condition also remarked \textit{``Every time I've typed my password in the PIN pad, I've said it out loud a few times to try to remember it.''}

In contrast to the PIN-based passwords, most participants in the ADBA condition appeared to remember their passwords without any explicit strategy or effort. For example: \textit{``It was surprisingly easy to remember the password. I didn't have to think about it after the initial step!''}, \textit{``The image was near impossible to forget despite the fact that I barely tried to put to memory ... I felt significantly more confident I would remember the images over the numbers.''}.
%, and \textit{``I was surprised how well I remembered the password.''}. One participant noted that they intentionally chose directions without salient features to make the password harder to guess and used the digits to facilitate memorization of those directions.

Beyond discussions of memorization strategies, three participants noted that selecting small targets on a PIN pad in VR using a laser pointer was not the most usable experience. For example: \textit{``It's a little bit hard to point to the pin pad sometimes''}, and \textit{``I feel the VR hand control method of entering the PIN was much harder than just a keypad for example.''}.%,  \textit{`It is like I had to recall the password and had to careful about clicking the correct number whereas for normal mouse/touchpad, I am only concern about the recalling part.''}.
% In addition to memorability, participants discussed
% - usability issue with controller pointing at small targets
% - study logistics (will be discussed later)
% - quest link connection issue
% - link cable length issue (will be discussed later)
% - security concerns (will be discussed separately)

\section{Discussion}

\subsection{Memorability}

ADBA demonstrated a clear memorability advantage over the PIN-based methods across both subjective and objective measures. This highlights that enforcing context-based password creation is key to achieving superior memorability, whereas in \cite{huang2024direction}, DBA did not show such an advantage as some participants relied on symbols during password setup for convenience.
%Notably, the actual difference may have been even greater than what was observed in the results. One participant who failed to recall their ADBA password after a week provided insightful feedback: \textit{``When making the password it would have been helpful after initially choosing it to go through it once highlighted with what you chose (like the recovery method) before confirming it three time.''}. This observation is particularly meaningful, as participants were allowed to review their chosen PINs after setup but were not afforded the same opportunity for ADBA. Consequently, this design element likely placed ADBA at a relative disadvantage in memorability. Nevertheless, even under these conditions, ADBA still outperformed the PIN-based methods.

The memorability advantages can be largely attributed to the well-established picture superiority effect. An additional contributing factor lies in ADBA's reduced interference between different environments, in contrast to the reuse of the same ten digits in PIN-based systems. We observed that many incorrect PIN entries resulted from confusion about ordering (e.g., attempting ``6-9-8'', ``6-8-9'', or ``8-6-9''). In contrast, ADBA users did not experience such confusion, as each of the six environments was unique.

More broadly, the reduced confusion in ADBA can extend from the positional to the password level. It is common for users to forget which PIN corresponds to which account. ADBA may mitigate this problem by using distinct environments for different authentications. Although certain visual similarities may persist across environments, they still preserve more differences than traditional PINs, where every position is drawn from the same ten digits.

A somewhat counterintuitive finding from our study was that the 6-digit PIN condition did not yield better memorability than the Random-PIN condition. A likely explanation is that participants in the 6-digit PIN condition often chose weak, patterned combinations that they assumed would be easy to recall. In contrast, participants assigned to the Random-PIN condition appeared more deliberate and attentive in memorizing their passwords.

% ADBA showed clear advantage in memorability.
% Mention a common source of error being confusion about the order
% ADBA was put at disadvantage by study design because participants did not have the chance to review their selections.

\subsection{Efficiency}
\label{sec:disc-efficiency}

Throughout the study, participants required more time to correctly enter a password using ADBA compared to the PIN-based methods. However, in contrast to this objective inefficiency, participants rated ADBA as even more efficient than the 6-Digit-PIN. This seemingly paradoxical result carries several important implications.

First, shorter entry time does not necessarily translate to more efficient authentication. Most prior authentication studies have operationalized efficiency through repeated entry of given passwords, and use the entry time as a proxy of efficiency~\cite{george2019investigating,mathis2021fast,george2017seamless,huang2024direction,khamis2018vrpursuits}. While such tasks reasonably estimate the lower bound of input time for a given method, they do not always reflect practical efficiency during real-world authentication scenarios. 
%For instance, if participants were asked to enter the same password repeatedly as in prior studies, ADBA would likely appear least efficient if entries were made via context, as the per-entry duration was substantially longer than that of the PIN-based methods, and the total task completion time would be scaled by the number of repetition. However, 
In our study, participants performed full authentication tasks rather than repeated entries. Many participants using ADBA were able to authenticate successfully on the first attempt without deliberate recall. In contrast, those in the PIN-based conditions often made several incorrect attempts, potentially spending additional time for recall and only then succeeded. Thus, although the per-entry time of ADBA was longer, the total authentication duration including recall and unsuccessful attempts was not necessarily longer. 
\textcolor{red}{Unfortunately, similar to prior work, we recorded only the time required for each individual entry and did not measure the time participants spent between entries. Therefore, the results support this interpretation as a plausible explanation, but do not provide conclusive evidence that it reflects the overall authentication time. We encourage future studies to measure the full authentication time, including the intervals between entries, rather than isolated entry times.}

%On the other hand, if participants were performing repeated entries the efficiency burden would be elevated whereas memorability burden reduced, so they would likely have switched to using digits,

%In addition, to discourage cheating, participants who forgot their passwords were simply reminded of them. In real-world contexts, however, users typically face account lockouts after several failed attempts and must reset their passwords if forgotten. Consequently, the true cost of efficiency due to poor memorability is much higher than what was reflected in our experiment.

%We hypothesized that participants in the ADBA condition might switch to the alternative digit-based entry mode after becoming familiar with their passwords to improve efficiency, but this did not occur. 
%This finding is specific to the entry frequencies tested in this study and should not be interpreted as a design failure, as the high perceived efficiency ratings were the highest, and with higher usage frequency, users may have greater incentive to speed up entry by switching to digit-based input. However, the results indicated that even under the ``high-frequency” condition in this study, participants found the context-based entry in ADBA sufficiently efficient.

Additionally, contrary to our expectation and the common assumption that an authentication taking more than 20 seconds would be too slow for practical use, no participants found it necessary or preferable to switch to the faster alternative available, even under daily or every-other-day authentication. No participants even commented on efficiency in the open feedback, implying that efficiency was unlikely part of their concerns.
Of course, this finding may not generalize to all possible scenarios. Authentication at higher frequencies than considered in this study, such as multiple times a day, may influence how users perceive efficiency, but the overwhelming preference for the authentication method assumed to be slow under realistic authentication tasks highlights that focusing exclusively on benchmark values such as entry time may risks overlooking what users actually perceive and value for their experience. 

Overall, the results indicate that ADBA was successful in supporting users' efficiency needs under the authentication frequencies considered in the study.

%Overall, our findings highlight that perceived efficiency represents a critical yet often overlooked dimensions of authentication performance. Users appear to value the smoothness and effortlessness of the authentication process more than the absolute time required for a single successful entry. From a user-centered perspective, efficiency should be viewed as a convenience afforded to users, rather than a constraint imposed upon them.

% Contrary to what we expected, no one in the ADBA condition switched to the more efficient entry, yet they still consider ADBA to be highly efficient.
% This could be due to entry time != authentication time.
% Efficiency is a convenience for users, not a requirement on users.

\subsection{Security}
\label{sec:disc-security}
ADBA offers $10^6$ possible combinations, exceeding the commonly used 4-digit PIN. More importantly, the expanded password space does not compromise memorability, unlike the transition from 4- to 6-digit PINs~\cite{munyendo2022same}. In addition, because the digit representations of ADBA passwords are randomly generated, users are prevented from creating trivial or predictable digit sequences for convenience, a common issue with 6-digit PINs~\cite{markert2020pin,wang2017understanding}.

ADBA’s robustness against observation attacks is achieved through the randomized initial direction in each environment, which produces entirely different gestures and body movements even when entering the same password. \textcolor{red}{Observation resistance was not directly measured in this study. Therefore, no conclusions can be drawn regarding its performance. For reference, the DBA study [18] demonstrated robustness to observation attacks for all ADBA entry methods (symbol-only, context-only, and hybrid), albeit using a different set of environments, with each environment divided into eight rather than ten directions. Thus, the extent to which these findings generalize to ADBA remains an open question.}

Results from the present study, indicate that ADBA still exhibits a degree of password homogeneity in users' choices of context directions, which could increase the predictability of passwords. 
It should be noted that this homogeneity is highly environment-specific rather than being intrinsic to ADBA, so it should not be interpreted as ADBA being fundamentally insecure. Instead, it highlights that the security of ADBA relies on the careful selection of virtual environments to minimize homogeneity.
One possible approach is to conduct pilot studies to identify environments producing diverse user preferences. For example, in our study, environments 4 and 6 yielded relatively balanced directional choices and could serve as suitable candidates. Such pilot studies require minimal effort from participants, who only need to indicate their preferred direction in each environment. %Nonetheless, conducting these pilots may still introduce non-trivial overhead and may not always be feasible.
Another approach is to assign a unique set of environments to each user. For instance, AI-based generation could be employed to automatically create visually distinct environments with diverse directional features for each application or individual user. Even if certain directional biases persist at the population level, they would be far more difficult to exploit since each environment would be unique to a single user.

The vulnerabilities of digit-based PINs have been well-documented in prior literature~\cite{markert2020pin,wang2017understanding} and were also confirmed in our findings. Since existing passwords were not allowed, many participants tended to choose weak, easily memorable sequences. However, even with weak passwords, the memorability performance of 6-digit-PIN remained inferior to that of ADBA.
%
%It should also be noted that the two PIN-based condition in our study were using a fixed layout without randomization, so unlike the ADBA condition they were not expected to be robust against shoulder-surfing attack. If randomization were introduced to these conditions, the efficiency and memorability might have been worse than what were shown in the results, given that users would not have been able to locate the digits by muscle memory, or facilitate memorization with spatial information as discussed in section \ref{sec:qualitative}.

% Password homogeneity is still an issue, possible solutions include 1) mass piloting to select a small library of environments with diverse preferences 2) Minimize environment reuse with ai generated environments

\subsection{Usability and Task Load}
Despite no reliable difference in SUS scores among the three conditions, ADBA resulted in a lower perceived task load than the PIN-based methods, particularly compared to the 6-Digit-PIN. The NASA TLX subscale scores indicated differences primarily in physical demand, mental demand, and perceived performance.

Participants reported higher physical demand with ADBA because they preferred turning and searching for target, which required more physical movement than entering digits on a PIN-pad. %Although a radial PIN-pad was available in ADBA, no participant chose to use it under the study setting.
%Notably, two participants in the ADBA condition who reported relatively low SUS scores (both 72.5, near the lower quartile for ADBA) mentioned that their only difficulty stemmed from using a short link cable, which restricted head movement. This issue is not intrinsic to ADBA, as it can be used seamlessly on standalone or wireless headsets, or with longer cables. Nonetheless, these comments highlight that ADBA is less usable in setups where head rotation is physically constrained.

Participants reported lower mental demand with ADBA, primarily due to its superior memorability, with comments such as \textit{``near impossible to forget''} and \textit{``didn't have to think''}.
They also reported higher perceived performance with ADBA compared to 6-Digit-PIN, though no reliable difference was found between ADBA and Random-PIN. The higher perceived performance relative to 6-Digit-PIN can be attributed to participants’ higher recall rates and fewer \textcolor{red}{attempts required to successfully authenticate} with ADBA. The lack of difference between ADBA and Random-PIN may appear counterintuitive given the objectively poorer performance of the latter. A plausible explanation is that participants' perceived performance is relative to their expectation: participants were likely more confident in memorizing self-chosen 6-digit PINs, whereas those assigned random PINs had lower expectations, and thus felt they performed better than anticipated despite lower objective success.

Taken together, the results demonstrate clear advantages of ADBA over widely adopted PIN-based methods, including improved memorability and perceived efficiency, lower overall task load, and more secure digit representations. With appropriate virtual environment selection, ADBA represents a highly promising knowledge-based authentication approach for VR, and we encourage researchers and practitioners to further examine the applicability of this method.

\section{Limitations and Future Work}

\textcolor{red}{In this study, we prioritized comparing ADBA with widely adopted 6-digit PINs in ecologically valid settings because the primary goal was to evaluate its practical adoptability rather than to quantify its improvements over its predecessor. Because DBA was not included as a study condition for direct comparison, any improvements over DBA remain inferential rather than statistically validated.}

\textcolor{red}{While the remote design enhanced ecological validity, participants were unobserved during the study, which limited our control over the study procedure and the extent of the data that can be collected. This trade-off is worth noting in the study design.}

\textcolor{red}{The requirements on user-selected PINs in this study were imposed to ensure fair comparisons across conditions and protect user privacy. However, in some real-world scenarios, users may choose insecure PINs for memorability, thus the generalizability of our findings to real-world scenarios should be interpreted with caution.}

The choice of low and high frequencies in this study was constrained by logistics, as frequencies that are too high or too low would be difficult to sustain. In practice, authentication can occur at both higher (e.g., multiple times per day) and lower frequencies (e.g., weeks between logins) than those examined here, so the results may not generalize to all real-world usage scenarios. Building on these results, future work can target higher usage frequencies, under which users may be more incentivized to prioritize efficiency, to have a more thorough understanding of the performance and preferences between digit- and context-based entries.

Additionally, despite good coverage of participants across different age groups, the study had a limited sample size for detecting between-subject effects and was gender-imbalanced, which may affect the generalizability of the findings. Recruiting and balancing demographics for a remote, longitudinal study requiring personal VR access is challenging. Due to practical constraints, we focused on the minimum eligibility requirements such as access to VR headsets, and did not explicitly control for demographics like gender.

%Despite our emphasis of ecological validity by emulating real-world usage and allowing greater flexibility than pre-scheduled lab-based tasks, the study remained a controlled experiment. The interventions were still artificially introduced rather than embedded within an application participants would use organically. A field study with minimal procedural control would be needed to evaluate the performance and usability of ADBA in authentic contexts.
%

% we did not record full authentication time, only time per entry, which could have been informative. This was mentioned in section \ref{sec:efficiency}

Given the promising results, future research can investigate reliable methods for selecting or generating virtual environments that promote diverse preferences to mitigate password homogeneity, which is the primary limitation of ADBA identified in this study. Beyond the approaches discussed in Section~\ref{sec:disc-security}, theories and methods from visual attention research may also offer valuable insights.

Under the authentication settings considered in this study, users overwhelmingly prioritized memorability over sheer entry efficiency even under daily authentication. Future studies can explore whether higher authentication frequencies, such as multiple times a day, would affect user preferences, and deepen our understanding of the factors that shape authentication usability.

%Lastly, future studies can also examine how environment reuse influences the memorability and usability of ADBA. Reusing a set of validated environments across applications could reduce password homogeneity at minimal implementation cost. While such reuse may introduce interference between passwords, the interference is likely to remain lower than with traditional PINs, where every position draws from the same ten digits. 
%The main potential drawback of environment reuse lies in digit cueing: because the digit-to-direction mapping is randomized, users might encounter inconsistent associations across different passwords. However, given participants’ strong preference for context-based cues in this study, further research is warranted to understand the mechanisms of environmental cueing and its interaction with digit-based entry.

\section{Conclusion}
% Method
In this paper, we built upon Direction-Based Authentication and proposed Adaptive Direction-Based Authentication (ADBA) to decouple users' needs of memorability or efficiency temporally. ADBA requires passwords to be chosen based on environment contexts, but provides both the environments and a digit-based entry method during authentication.
% Study & Findings
The comparison between ADBA and the widely-used PIN-based methods over a longitudinal study with varying entry frequency periods demonstrated the superiority of ADBA in terms of better memorability, lower overall task load, and higher perceived efficiency even though the passwords took longer to enter compared to the PINs. ADBA also offered security benefits, alleviating memorability and usability issues associated with expanded password length for digit-based PINs, although depending on the environment used the choices of users may exhibit varying degree of homogeneity.
% Take-home message
The results of the study showed that ADBA is a promising user authentication technique for VR given the solid advantages over the widely-used PIN-based method, but future studies are warranted to explore and validate reliable environment selection methods to mitigate password homogeneity, and look further into users' prioritization of efficiency and memorability under different authentication frequencies and settings.

\bibliographystyle{ACM-Reference-Format}
\bibliography{sample-base}

\end{document}